# Nano watermill driven by the revolving charge

Xiaoyan Zhou,[1,2], Jianlong Kou,[2] Xuechuan Nie[3], Fengmin Wu[1,2,a)], Yang Liu[4,a)] and Hangjun Lu[2 a)]

[1]*Department of Physics and Institute of Theoretical Physics, Shanxi University, Taiyuan, 030006, China*

[2]*Department of Physics, Zhejiang Normal University, Jinhua, 321004, China*

[3]*Shanghai Institute of Applied Physics, Chinese Academy of Science, Shanghai, 201800,China*

[4]*Department of Mechanical Engineering, The Hong Kong Polytechnic University, Hong Kong, 999077, China*

Using molecular dynamics simulations, we propose a novel nanoscale watermill for unidirectional transport of water molecules through a curved single-walled carbon nanotube (SWNT). In this nanoscale system, a revolving charge is introduced to drive water chain confined inside the SWNT, which is served as nano waterwheel and nano engine. A resonance-like phenomenon is found that the revolving frequency of the charge plays a key role in pumping water chain. The water flux across the SWNT increases with respect to the revolving frequency of the external charge and reaches the maximum when the frequency is 4 THz. Correspondingly, the number of the hydrogen bonds of water chain inside the SWNT decreases dramatically with the frequency ranging from 4 THz to 25 THz. The mechanism behind the resonant phenomenon has been investigated systematically. Our findings are helpful for designing nanoscale fluidic devices and energy converters.

a) Electronic mail:wfm@zjnu.cn or yang.liu@polyu.edu.hk or zjlhjun@zjnu.cn.

# I. INTRODUCTION:

Actuation of a fluid flow in the nanopores is of fundamental importance for the progress in design and utilization of novel nanofluidic devices, machines and sensors, which have a broad prospect of industrial applications including nanofiltration, water purification, and hydroelectric power generation.[1-6] In fact, not only our daily life but also life itself depends on the transport of water through pipes, capillaries, and protein channels, controlled by pumps, valves, and gates.[7-12] Fluid pumping is an essential function of a nanofluidic system.[13] It has been recognized that the active transport of water through nanopores is technically difficult or even impossible using conventional methods due to the large surface to volume ratio, though there are various kinds of devices applied to pump water conveniently on macroscopic level. Therefore, it is crucial to develop effective water pump devices which can make a continuous unidirectional water flow on nanoscale.

Over the last two decades, the field of nanofluidic has seen rapid development. Various novel concepts and blueprints for nanoscale pumps based on the carbon nanotubes have been proposed,[3, 14-25] owing to a number of attractive features of the carbon nanotubes for fabrication of nanofluidic devices.[26-28] In 1999, Král *et al.* proposed a laser-driven pump for atomic transport through carbon nanotubes.[29] Laser was applied to excite an electric current in the carbon nanotube, which drives intercalated atoms by the wind force. The experiments that followed have demonstrated that nanoparticles of iron inside the carbon nanotubes can be driven in the direction of the electron flow by electromigration force.[30] By using molecular dynamics simulation, Insepov *et al.* found that the gas inside the carbon nanotube can be driven by the Rayleigh surface wave.[3] Nanoscopic propellers, designed by robust macroscopic principles and possessing "chemically tunable" blades, were introduced by Wang and Král to pump solvent molecules.[31] In 2012, a rotary nano ion pump, inspired by the $F_0$ part of $F_0F_1$-ATP synthase biomolecular motor was investigated by molecular dynamics simulation.[32] Simulation results demonstrated that an ion gradient would be generated between the two sides of the nanopump when the rotor of the nanopump rotated mechanically. Duan *et al.* used a small portion of the initially twisted wall of a carbon nanotube to function as an energy pump for transportation of water molecules.[19] Using molecular dynamics simulations, Chang demonstrated that the domino wave along the longitudinal direction of the tube can be developed. The molecules inside the SWNT can be pumped by the domino wave.[33] Ma's group demonstrated that water can be pumped by revolving chiral carbon nanotube.[21] In addition, progress in moving water has been made by designing systems with an imbalance of surface tension or a chemical or thermal gradient,[20, 34-36] but it is still difficult to make a controllable continuous unidirectional water flow.[18]

Recently, various nanopumps driven by electric fields or electric current have been proposed.[14-16, 37-39] Král and Shapiro[40] predicted that electric current can be generated in metallic carbon nanotubes immersed in liquids flowing along them, which was consistent with the following experimental results of Ghosh *et al.* two years later.[41] Inspired by this result, Sun *et* *al.* demonstrated experimentally that a water flow can also be driven by the applied current of an SWNT.[1] In our previous works, we have proposed nano water pumps based on ratchet-like mechanism without osmotic pressure or hydrostatic drop.[23, 35, 42]

In this paper, we propose a novel blueprint for nano watermill by numerical simulations, where a revolving charge serves as waterwheel. Experimentally, the revolving charge can be achieved by an array of electrodes with certain mode or by some molecular rotors.[43, 44] It is noted that molecular rotors with a fixed off-center rotation axis have been observed recently by a scanning tunneling microscope.[45] Moreover, the fabrication methods of nanofluidic channels and nanotubes are developed.[46, 47] In fact, our idea is inspired by these experimental works. We found that the continuous directional water flux can be pumped effectively by a revolving charge. The revolving frequency of the external charge plays a key role in water transportation. An interesting resonance was found that the water flux peaks at $f$ = 4 THz. However, the water chain confined inside the SWNT is disturbed and the hydrogen bonds between them were ruptured. Therefore, the number of the hydrogen bonds decreases dramatically. The mechanism behind the resonant phenomenon has been investigated systematically. Our simulation results demonstrate that the nano watermill can convert the energy of the revolving charge into the transport of the water molecules inside the curved carbon nanotube conveniently.

## II. METHODOLOGY

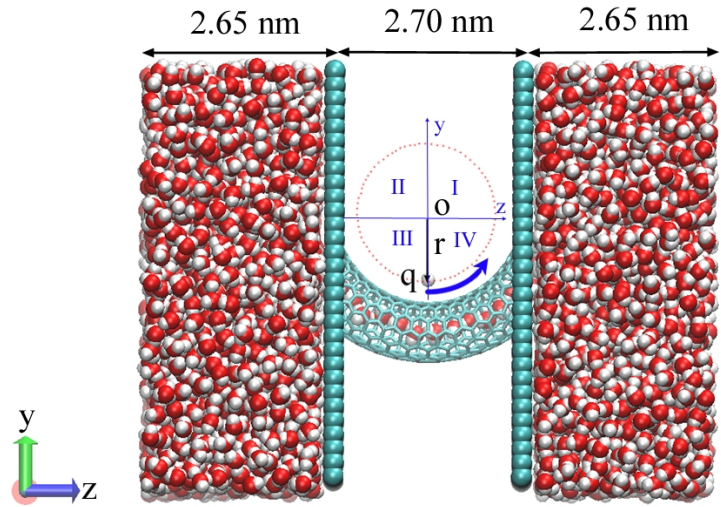

Fig. 1. (Color online) Sectional view of the simulation framework: the charge (white ball) $q$ rotates around the origin O with radius $r$ = 1.0 nm. The carbon nanotube with 8.1 Å diameter and two carbon sheets are shown in light blue. Water molecules are depicted in VDW representation (image created with VMD [1.8.7]).

The simulation framework is illustrated in Fig. 1. Membranes were created by two carbon sheets with the distance of 2.7 nm. A curved (6, 6) SWNT with 8.1 Å diameter was embedded in the membranes along the $z$ direction, as shown in Fig. 1. The curvature radius of the SWNT is 1.755 nm. A positive charge with a quantity of $q$ (= +1$e$) is introduced outside the tube. It rotates in a circle of radius $r$ (=1.0 nm) with a constant angular velocity $\omega$. The radial distance of the circle point O from the carbon tube wall is 1.35 nm. To keep the simulation system electronically neutral, a contrary charge -$q$ is assigned close to the boundary.

MD simulations were performed with GROMACS 4.0.7 [48, 49] at a constant volume with the initial box size dimensions of $L_x$ = 6 nm, $L_y$ = 6 nm, $L_z$ = 8 nm, and a temperature of 300 K for 100 ns. Periodic boundary conditions were applied in all directions. The simulation box contains 8317 water molecules which are modeled by using the TIP3P model.[50] A time step of 2 fs was adopted when the revolving frequency is less than 10 THz and 1fs was adopted for others. Simulation data were collected every 1 ps. The last 95 ns were collected for analysis. The carbon nanotube is regarded as a large molecule. The total potential energy function of CNT can be expressed in the form:

$$V_{total} = \sum_{VDW} 4\varepsilon[(\frac{\sigma}{r})^{12} - (\frac{\sigma}{r})^{6}] + \sum_{bond} \frac{1}{2}k_b(r-r_0)^2 + \sum_{angles} \frac{1}{2}k_\theta(\theta-\theta_0)^2 + \sum_{dihedrals} \frac{1}{2}k_\xi(\xi-\xi_0)^2 .$$

where $\sigma_{CC}$ = 0.34 nm, $\varepsilon_{CC}$ = 0.3612 kJ mol$^{-1}$, bond lengths of $r_0$ = 0.142 nm, bond angles of $\theta_0$ = 120°, spring constant of $k_b$ = 393 960 kJ mol$^{-1}$ nm$^{-2}$ and $k_\theta$ = 527 kJ mol$^{-1}$ deg$^{-2}$, $k_\xi$ = 52.718 kJ mol$^{-1}$ deg$^{-2}$. The Lennard-Jones parameters for the interaction between a carbon atom and the water oxygen are $\varepsilon_{CO}$ = 0.4802 kJ mol$^{-1}$ and $\sigma_{CO}$ = 0.3275 nm. With periodic boundary conditions, long range electrostatic interactions with a cutoff for real space of 1.4 nm were computed by using a particle-mesh Ewald method. In each simulation, the carbon sheets are fixed.

### III. RESULTS AND DISCUSSION

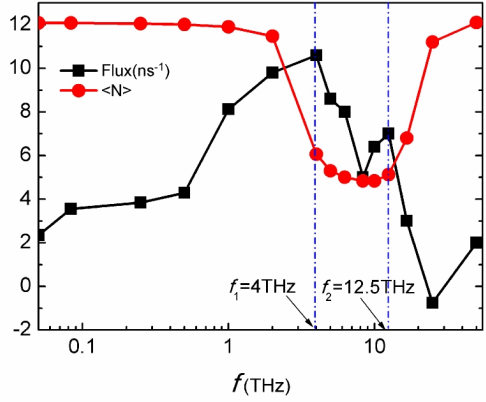

Fig. 2. (Color online) Water flux and average number of water molecules N inside the curved SWNT as a function of the revolving frequency $f$ of the external charge.

In order to study the directional transport of the water through the curved SWNT driven by a revolving charge, we define flux as the difference between the number of water molecules leaving from the left end and the right end (again having entered from the opposite end) per nanosecond.

Fig. 2 shows water flux and the average number of water molecules inside the curved nanotube as a function of the revolving frequency of the charge. For $f$ = 50 GHz, the curved SWNT is occupied by about 12 water molecules, which is determined by local excess chemical potential [26]. Water molecules not only penetrate into, but are also conducted through, the SWNT. The water flux is about 2.35 water molecules per nanosecond through the nanotube. This value is comparable to the measured 3.9 ns$^{-1}$ for aquaporin-1.[51] From the Fig. 2, we can see that the net flux is very sensitive to the revolving frequency $f$. It increases remarkably to 10.6 ns$^{-1}$ when the frequency $f$ increases from about 0.05 THz to 4 THz. This maximum value is about four times the water flux (about 2.35) at $f$ = 0.05 THz. Interestingly, there are two peaks at $f_1$ = 4 THz and $f_2$ = 12.5 THz. Correspondingly, the average number of the water molecules inside the curved SWNT decreases dramatically to about 5, indicating that the single file of water chain is disturbed dramatically. The net flux decreases sharply when the frequency is larger than $f_2$. Surprisingly, the average number of water molecules inside the curved SWNT recovers to 12, indicating that the curved SWNT is filled by water molecules again. Moreover, the direction of the flux even reverses at $f$ = 25 THz. The critical frequency of the single-file water chain can be determined by using the relation $E_{HB} = m\omega^2 A^2/2$, where $E_{HB}$ is the binding energy of the hydrogen bond, $m$ is the mass of water molecule, and $A$ is the amplitude of allowed radial motion. The value of $E_{HB}$ is about 16 kcal/mol and the amplitude is about 0.3-0.9 Å[53]. The classical resonant frequency of the water chain inside the (6, 6) SWNT is estimated to lie between 4.8 THz and 14.5 THz. Our simulations results show that the water flux peaks at $f_1 = 4$ THz and $f_2 = 12.5$ THz which is well accordant with the theoretical values and results in the previous works.[52,]

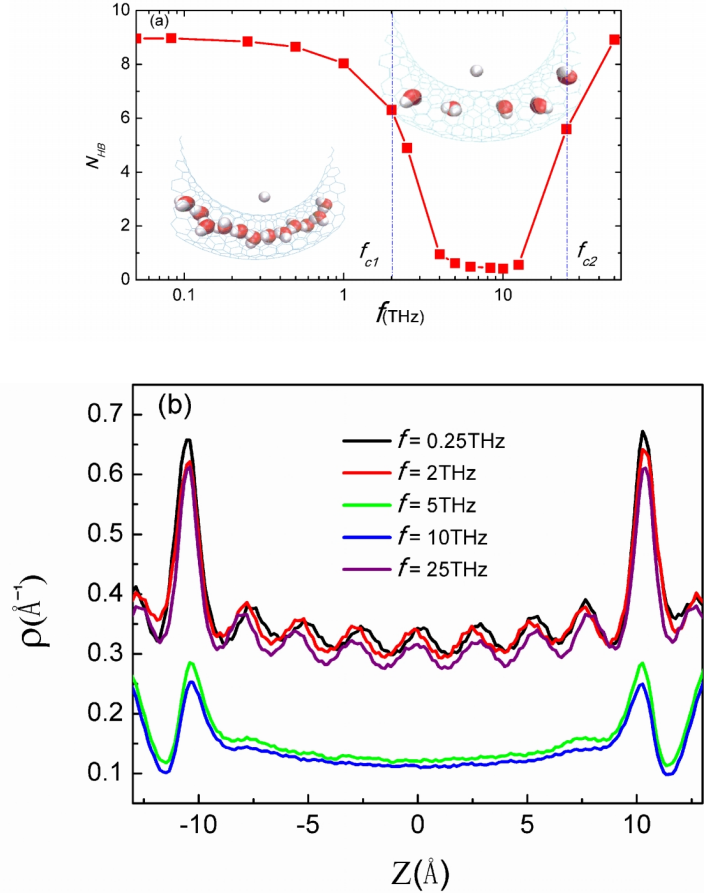

Fig. 3. (Color online) (a) Average number of the hydrogen bonds inside the curved SWNT for different rotation frequency. (b) Water density distributions inside the nanotube along $z$ axis under different rotation frequency.

Water molecules are connected by hydrogen bonds inside the SWNT, which are shielded from fluctuations in the bulk water. Hydrogen bonds are really a special case of dipole forces,[54] and play a key role in the molecular transportation through the nanotube. Here, we focus on the hydrogen bonds connecting between the water molecules inside the SWNT. And a geometric definition of hydrogen bonds is adopted. Water pair is hydrogen-bonded if the O–O distance was less than 3.5 Å and simultaneously the bonded O–H···O angle was less than 30°.

The average number of the hydrogen bonds (shown in Fig. 3 (a)), denoted by $N_{HB}$, shares the same trend as the average number of water molecules inside the tube (shown in Fig. 2), it decreases very slowly for $f < f_{c1} = 2$ THz. The average number of water molecules is about 12 and the average number of hydrogen bonds is about 9 in the range of $f < f_{c1}$. The average number of water molecules and hydrogen bonds are both almost unchanged, indicating that the single-file water chain inside the SWNT is affected slightly by the external revolving charge. Furthermore, from Fig. 3 (b), we can see the wavelike pattern of the water density distribution along the z direction. The wavelike pattern is sharp near the ends of the curved SWNT. However, when $f_{c1}$ (=2 THz) $< f < f_{c2}$ (= 25 THz), the average number of water molecules decreases dramatically to about 5. Correspondingly, the average number of hydrogen bonds, $N_{HB}$, decreases sharply to 0.5. All of the hydrogen bonds are almost disrupted by the revolving charge. The water molecules transport across the nanotube individually, not collectively. As shown in Fig. 3 (b), the wavelike pattern of water density distribution is quite flat near the center of the nanotube. Surprisingly, when the frequency is larger than 25 THz, $N_{HB}$ increases with the frequency. In the range of $f > f_{c2}$, $N_{HB}$ increases sharply to 9. The wavelike pattern of water density distribution becomes obvious again. The influence of the charge rotation on the structure of the water chain becomes negligible.

Formations and ruptures of hydrogen bonds occur not only in water, but in some alcohols and their aqueous solutions. We have also conducted simulations to study the effect of revolving charge on behavior of methanol confined inside the carbon nanotube. We found that there was similar resonance phenomenon. However, a downshift of the critical frequencies for the methanol system compared to that of water system is found. It is mainly due to their different strength of the hydrogen bonds and the molecular mass. Water is a strongly polar molecule while methanol is intermediate between nonpolar and strongly polar molecules.[55]

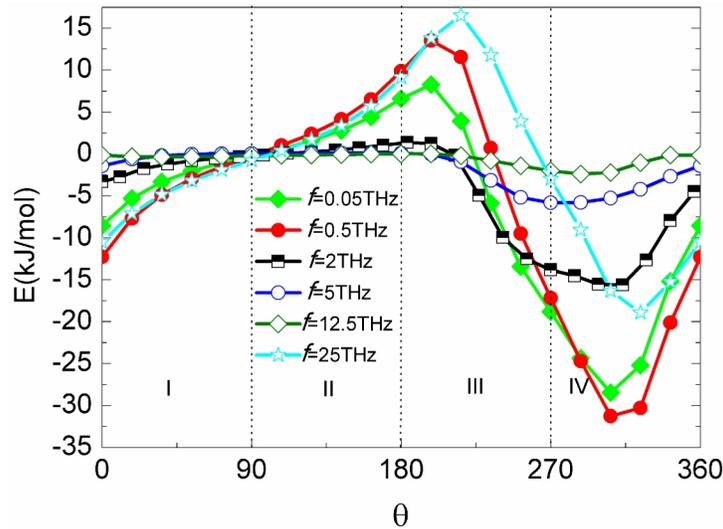

Fig. 4. (Color online) The interaction energies between the revolving charge and the water molecules inside the curved SWNT for different frequency.

To understand the mechanism behind the nano water pump, we calculated the water charge interaction energies for different frequency of the revolving charge. Because of the external charge rotates around the center point O, the position of the charge can be depicted by the angle $\theta$ conveniently, as shown in Fig. 1. And $\theta$ is the angle between the position vector **r** of the revolving charge and z-axis. The size of the angle, $\theta$, is measured in degrees. At time $t = 0$, the point charge is on the reference line (*z* axis). The axes (*y* and *z*) of a two-dimensional Cartesian system divide the plane into four regions, called quadrants. When $0 < \theta < 90°$, the point charge locates in the 1st quadrant, denoted by Ⅰ. Similarly, the other three quadrants denoted by Ⅱ, Ⅲ, Ⅳ, respectively.(see Fig. 4).

From Fig. 4, we can find that the profile of interaction energy is asymmetrical. When the revolving charge moves in the quadrant Ⅰ, the interaction energies are negative, indicating that the charge drags water molecules inside the SWNT. The attractive energies decrease with $\theta$ due to increase of the distance between the charge and water molecules. The distance reaches maximum when $\theta$ is 90°, resulting in the charge water interaction energies decrease to about zero. On the contrary, interaction energies increase and become positive in the quadrant Ⅱ, indicating that the revolving charge pushes water molecules through the carbon nanotube. Furthermore, $180° < \theta < 225°$, the majority of the interaction energies are positive, due to the dipoles of the water molecules inside the SWNT cannot flip synchronously. When $\theta > 225°$, the interaction energies between the charge and water molecules inside the SWNT become negative. However, the peaks don't locate in $\theta = 270°$, in which the average distance between the charge and the water molecules inside the SWNT decreases to the smallest. So, the pumping effect is attributed to the asymmetry interaction energy acting on the water molecules inside the nanotube.

The profiles of the interaction energies depend on the revolving frequency of the charge. When $f < f_{c1}$ or $f > f_{c2}$, the interaction energies vary obviously with the $\theta$ (or the position of the point charge). In the range of $f_{c1} < f < f_{c2}$, the interaction energies becomes weak due to the average number of water molecules inside the SWNT decreases dramatically.

## IV. CONCLUSIONS

In the present work, we design a novel nanopump with a SWNT and a revolving charge outside. We have shown the excellent pumping effect of this water pump. The net flux increases rapidly with the frequency of the revolving charge. An interesting resonance phenomenon is found. As $f = 4.0$ THz, the net flux reaches 10.6 ns$^{-1}$, which is about three times the measure 3.9 ns$^{-1}$ for AQP. We find that the remarkable pumping effect is attributed to the asymmetry interaction energy acting on the water-chain inside the nanochannel. When the revolving charge moves into the quadrants Ⅲ and Ⅳ, the interaction acting on the water chain is strong enough to drag the water chain across the right outlet of the SWNT. Then the charge moves into the regions Ⅰ and Ⅱ, the interaction is so weak that it is difficult to push the water chain towards the left entrance. Interestingly, we found that the single-file water chain remains well even the frequency of the charge rotation reaches a large value of 1.5THz. Our design is expected to have some implications in nanotechnology, such as hydroelectric power converters, fluid separation, drug delivery and sensor applications.

**ACKNOWLEDGMENTS**

We thank Haiping Fang and Ruhong Zhou for helpful discussions. This work is partially supported by the National Natural Science Foundation of China (Grant Nos. 11005093 and 61274099), Research Fund of Department of Zhejiang Provincial Education under Grant No. Y201223336, Zhejiang Provincial Science and Technology Key Innovation Team No. 2011R50012 and Zhejiang Provincial Key Laboratory No. 2013E10022, and The Hong Kong Polytechnic University (G-YL41).